\documentclass[aps,prl,twocolumn,float]{revtex4}
\usepackage{amsmath}
\usepackage{pstricks}
\usepackage{pst-node}
\usepackage[ansinew]{inputenc}
\usepackage{amssymb,amsmath}
\usepackage{amsfonts}
\usepackage{epsfig}
\usepackage[english]{babel}
\usepackage{graphics}
\usepackage{hyperref}

\def\g{\gamma}

\def\o{\omega}

\def\e{\varepsilon}

\font\Sets=msbm10
\def\Integer {\hbox{\Sets Z}}

\def\Natural {\hbox{\Sets N}}

\def\be{\begin{equation}}
\def\ba{\begin{array}}
\def\ee{\end{equation}}
\def\ea{\end{array}}
\def\bea {\begin{eqnarray}}
\def\eea {\end{eqnarray}}
\def\bean{\begin{eqnarray*}}
\def\eean{\end{eqnarray*}}

\def\RA {\ \Rightarrow\ }

\begin{document}

\title{Resonant interactions of nonlinear water waves in a finite basin}
\author{Elena Kartashova$^{\dag\ddag}$, Sergey Nazarenko$^{}$, Oleksii Rudenko$^{\dag}$\\
$^{\dag}$Weizmann Institute of Science, Rehovot, Israel\\
$^*$Mathematics Institute, University of Warwick, Coventry CV4-7AL,
UK\\
$^{\ddag}$RISC, J. Kepler University, Linz, Austria\\
e-mail: lena@risc.uni-linz.ac.at}

\begin{abstract}
We study exact four-wave resonances among gravity water waves in a
square box with periodic boundary conditions. We show that these
resonant quartets are linked with each other by shared Fourier modes
in such a way that they form independent clusters. These clusters
can be formed by two types of quartets: (1) {\it angle-resonances}
 which cannot directly cascade energy but which can
redistribute it among the initially excited modes and (2) {\it
scale-resonances} which are much more rare but which are the only
ones that can transfer energy between different scales. We find such
resonant quartets and their clusters numerically on the set of 1000
x 1000 modes, classify and quantify them and discuss consequences of
the obtained cluster structure for the wavefield evolution. Finite
box effects and associated resonant interaction among discrete wave
modes appear to be important in most numerical and laboratory
experiments on the deep water gravity waves, and our work is aimed
at aiding the interpretation of the experimental and numerical data.
\end{abstract}

 \maketitle

\tableofcontents
%\newpage
\setlength{\parindent}{0pt} \setlength{\parskip}{\medskipamount}

%-------------------------------------------------------------------------

\section{Introduction}
Weakly nonlinear systems of random waves are usually studied
in the framework of wave turbulence theory (WTT) (for introduction to WTT see book
 \cite{lvov}).
 Besides weak nonlinearity and  phase randomness,
 %(equivalently, fast decay of spatial correlations)
  this statistical
 description is based on the infinite-box limit.
 This approach  yields wave kinetic equations for the wave spectrum
 which have important
stationary solutions, Kolmogorov-Zakharov  (KZ) spectra \cite{lvov}.
Importance of KZ spectra is in that they correspond to a constant
flux of energy in Fourier space and, therefore, they are analogous
to the Kolmogorov energy-cascade spectrum in hydrodynamic
turbulence. WTT approach can also  be used to study
 evolution of higher momenta of wave amplitudes
and even their probability density function, and, therefore,  to
examine conditions for deviation from Gaussianity and onset of
intermittency \cite{nnb01,ln04,clnp04,cln04}. Significant effort has
been done in the past to test WTT and its predictions numerically
\cite{clnp04,LNP-2006,z1,z2,shrira} as well as experimentally
 \cite{denis,fauve,kolm,braz}. It has been noted however that in both the numerical
 and the laboratory experiments the domain boundaries typically play
 a very important role and the infinite-box limit assumed by WTT is
 not achieved. Indeed, as mentioned in \cite{LNP-2006}, to overcome the
wavenumber discreteness associated with the final box one needs at
least 10000 x 10000 numerical resolution which is presently
unavailable for this type of problems. On the other hand, as shown
in \cite{denis}, even in a 10m x 6m laboratory flume the finite-box
effects are very strong. It is important to understand that WTT
takes the infinite-box
 limit before the weak-nonlinearity limit, which physically means that
 a lot of modes interact simultaneously if they are in {\em quasi-resonance}, i.e.
satisfy the following conditions
 \bea\label{res}
\begin{cases}
|\omega ({\bf k}_1) \pm \omega ({\bf k}_2)\pm ... \pm \omega ({\bf k}_{s})| < \Omega,\\
{\bf k}_1 \pm {\bf k}_2 \pm ... \pm {\bf k}_{s} = 0
\end{cases}
\eea %%
with some resonance broadening $ \Omega>0$ which is a monotonically
increasing function of nonlinearity (mean wave amplitude). Here $
{\bf k} $ and $\o({\bf k}) $ are wave vector and dispersion function
(frequency) which correspond to a general wave form $\sim \ \exp{i(
{\bf k}{\bf x} -\omega t)}.\ $ WTT is supposed to work when the
resonance broadening $\Omega$ is greater than the spacing
$\delta_\omega$ between the adjacent wave modes \be
 \Omega > (\partial \omega
/\partial k) 2 \pi/ L,
\label{quasi-r}
\ee
 where $L$ is the box size.
 For some types of waves, for example for the capillary water waves,
 this condition is easy to satisfy, and therefore to achieve WTT regime
 \cite{alstrom}.
However, this condition is often violated for some other types of
waves, in particular for the surface gravity waves which will be the
main object of this paper. This occurs in numerical simulations, due
to limitations on the numerical resolution \cite{LNP-2006,z2} and in
laboratory experiments, due to an insufficient basin size
\cite{denis,fauve}. In these cases, the Fourier space discreteness
(which is due to a finite box size) leads to significant depletion
of the number of wave resonances with respect to the infinite box
limit. In turn, this results in a slowdown of the energy cascade
through the $k$-space with respective steepening of the wave spectra
\cite{Naz-2006,denis}. In addition, the wavenumber grid will cause
the wave spectra to be anisotropic in this case.

What happens when the condition (\ref{quasi-r}) is so badly violated
that only waves which are in exact resonance (i.e. $\Omega =0$) can
interact? In this case, the mechanism of the wave phase randomization
based on many quasi-resonant waves interacting simultaneously
will be absent and, therefore, one should expect less random
and more coherent behavior.
In \cite{PRL94} it was shown that in many wave
systems, resonantly interacting waves in finite domains
 are partitioned into small
independent clusters in Fourier space, such that there cannot be an
energy flux between different clusters. In particular, in \cite{AMS}
some examples of wave systems were given in which no resonances
exist (capillary water waves, $\o=|{\bf k}|^{3/2}$), as well as
systems with an infinite number of resonances  (oceanic planetary
waves, $\o=|{\bf k}|^{-1}$). Both of these examples are three-wave
systems [i.e. $s=3$ in (\ref{res})]. In the present paper we will
concentrate on finding resonances for the deep water gravity waves,
which is a four-wave system.

The problem of computing exact resonances in a confined laboratory
experiment is highly non-trivial because the wavenumbers are
integers and (\ref{res}) is a system of Diophantine equations on
many integer variables in large powers. Computational time of
solving this system by a simple enumeration of possibilities in this
case grows exponentially with each variable and the size of spectral
domain under consideration. A specially developed $q$-class method
\cite{K06-3, KK06-1, KK06-2} has allowed to accelerate the
computation and find all the resonances among  waves in large
spectral domains, in a matter of minutes. In this paper we use
$q$-class method to construct resonant wave clusters formed by
4-wave resonances among water gravity waves covered by the kinematic
resonance conditions in the form: %%
 \bea\label{4grav}
\begin{cases}
|{\bf k}_1|^{1/2} + |{\bf k}_2|^{1/2}= |{\bf k}_3|^{1/2} + |{\bf k}_4|^{1/2},\\
{\bf k}_1 + {\bf k}_2 = {\bf k}_{3} + {\bf k}_{4}
\end{cases}
\eea %%
with ${\bf k}_i = (m_i, n_i)$ and integer $|m_i|, |n_i| \le 1000.$
Our main aim is to understand how anisotropic resonance clusters
influence the general dynamics of the complete wave field.

\section{Construction of $q$-classes}
We adopt the general definition of a $q$-class given in \cite{K06-3} for
the dispersion function $\o=|{\bf k}|^{1/2}$ in the following way. Consider
the set of algebraic numbers $R=±k^{1/4}$. Any such number $k$ has a
unique representation
$$
    k = \g q^{1/4} , \g \in \Integer ,
$$
where $q$ is a product
$$
    q=p_1^{e_1} p_2^{e_2} ... p_n^{e_n},
$$
while $p_1, ... p_n$ are all different primes and the powers $e_1,
....e_n \in \Natural $ are all smaller than $4$. Then
    the set of numbers from $ \mathbb{R}$ having the same $q$ is called a $q$-class $Cl_q$.
    The number $q$ is called a class index. For a number $k = \g q^{1/4}$,
    $\g$ is called the weight of $k$. For instance,
 wave vector ${\bf k}=(160,
 40)$ belongs to the $q$-class with $q=1700.$ Obviously, for any two numbers $k_1, k_2$ belonging to the same
$q$-class, all their linear combinations with integer coefficients
 belong to the same class $q$.

It can be shown that Sys.(\ref{4grav}) have two general types of
solutions:

{\it Type I:} All 4 wave vectors belong to the same class $Cl_{q},$
in which case first equation of the Sys.(\ref{4grav}) can be
rewritten as%%
\be\label{1class} \g_1\sqrt[4]{q} +
\g_2\sqrt[4]{q}=\g_3\sqrt[4]{q}+\g_4\sqrt[4]{q} \ee%%
with integer $\g_1, \g_2, \g_3, \g_4.$

and

{\it Type II:}  All 4 wave vectors belong to two different classes
$Cl_{q_1}, Cl_{q_2};$ in this case first equation of the
Sys.(\ref{4grav}) can be rewritten as %%
\be\label{2class} \g_1\sqrt[4]{q_1} +
\g_2\sqrt[4]{q_2}=\g_1\sqrt[4]{q_1}+\g_2\sqrt[4]{q_2} \ee%%
with integer $\g_1, \g_2.$

Notice that (\ref{2class}) {\it is not an identity} in the initial
variables $m_i, \ n_i$ because an integer can have several different
presentations as a sum of two squares, for instance, 4-tuple
\{\{-1,4\},\{2,-5\},\{-4,1\},\{5,-2\}\} is an example of II-type
solution, with all weights $\g_i=1$ and $q$-class indexes $q_1=17,
q_2=29$.

The I- and II-type of solutions describe substantially different
energy exchanges in the $k$-space. The II-type resonances are called
{\it angle-resonances} \cite{K07} and
 consist of wavevectors with pairwise equal lengths,
 i.e.  $|{\bf k}_1| = |{\bf k}_3|$ and  $|{\bf k}_2| = |{\bf k}_4|$
 or $|{\bf k}_1| = |{\bf k}_4|$ and  $|{\bf k}_2| = |{\bf k}_3|$.
 Thus, these resonances do not transfer energy outside of the
 initial range of $|{\bf k}|$ and, therefore, cannot provide
 an energy cascade mechanism. However, these resonances can redistribute
 energy among the initial wavenumbers, in both the direction ${\bf k}/|{\bf k}|$
 and the scale $|{\bf k}|$. Since the initial support of energy
 in $|{\bf k}|$ cannot change, the II-type resonances alone
 would form a finite dimensional system, and it would be reasonable expect
 a relaxation of such a system to a thermodynamic Rayleigh-Jeans distribution
 determined by the initial values of the motion integrals (the energy and the waveaction
 in the this case). Note that such a thermalization
 could happen only among the resonant wavenumbers, and
 many modes which are initially excited but not in resonance would not
 evolve at all. (Such an absence of the evolution was called "frozen turbulence"
 in \cite{z1} where the capillary waves were studied for which there are no exact resonances).
 Whether the thermalization does occur
in finite clusters and under what conditions (e.g. the cluster size
etc.) are interesting questions that remain to be studied in future.

 On the other hand, the I-type resonances are called
{\it scale-resonances} \cite{K07},  and they can generate new
wavelengths, - this corresponds to 3 or 4 different weights $\g$ in
(\ref{1class}) \cite{K07}. Thus, they are the only kind of
resonances that can transfer energy outside of the range of initial
$|{\bf k}|$.

\begin{figure}[h]
  \begin{center}
    \includegraphics[width=4cm]{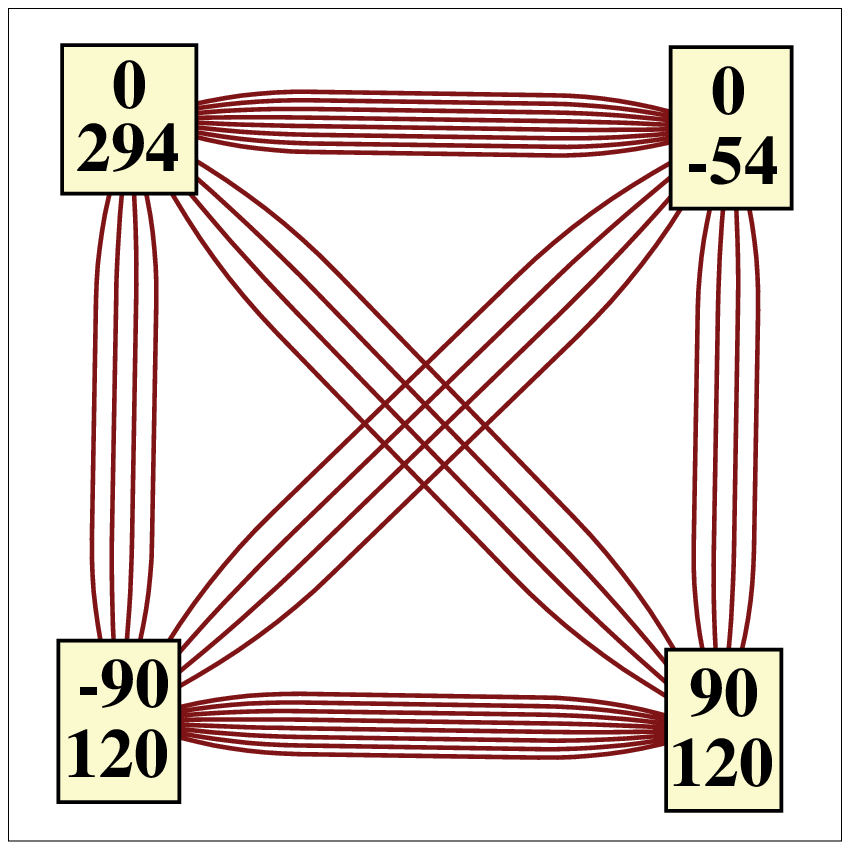}
    \includegraphics[width=4cm]{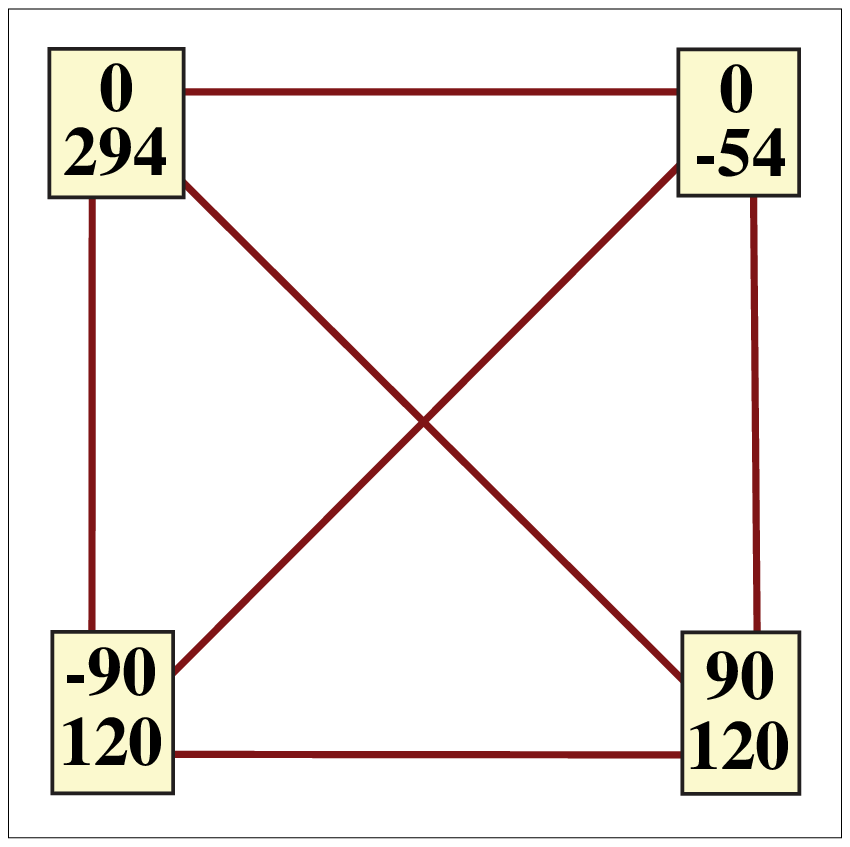}
  \end{center}
\caption{\label{f:smallest} Color on line. The smallest tridents
cluster formed by 8 symmetry generated solutions, two presentation
forms are given: with multi-edges (left panel) and without
multi-edges (right panel)}
\end{figure}

 Before studying the structure of resonances let us notice the
following simple but important fact.
Suppose a quartet %
\be\label{sym}  \{\tilde{\bf k}_1,\tilde{\bf
k}_2,\tilde{\bf k}_3,\tilde{\bf k}_4\}\ee%
 is a solution of
(\ref{4grav}), then each permutation of indexes $1 \leftrightarrow
2, 3\leftrightarrow 4 $ or simultaneous $1 \leftrightarrow 3 $ and $
2 \leftrightarrow 3$  will generate a new solution of (\ref{sym});
in general case, all together 8 different {\it symmetry generated
solutions}. Of course, in some particular cases, when some of the
vectors belonging to a quartet coincide, the overall number of
symmetry generated solutions can be smaller. For instance, the
quartet $\{\{\{0,-54\},\{0,294\},\{90,120\},\{-90,120\}\}$ is part
of the cluster of 8 symmetry generated solutions, while the quartet
$\{\{17,31\},\{-153,-279\},\{-68,-124\},\{-68,-124\}\}$ belongs to
the cluster of 4 such solutions. On the Fig.~\ref{f:smallest} the
smallest tridents cluster is shown, with and without multi-edges.
Graphical presentation of a cluster as a graph with multi-edges is
of course mathematically correct but would make the pictures of
bigger clusters somewhat nebulous (see Fig.~\ref{f:smallest}, left
panel). Notice that although formally we have solutions with
multiplicities  due to the symmetries, physically all these
solutions correspond to the same quartet, and therefore we count
them as one. Therefore, further on we omit multi-edges in our
graphical presentations as it is shown in Fig.~\ref{f:smallest},
right panel.

%%

In the next sections, on all the Figures, the structure of resonance
clusters in the {\it spectral space} is shown as follows: each wave
vector is presented as a node of integer lattice and nodes belonging
to one solution are connected by lines.

\section{Scale-resonances}

We have studied the cluster structure of the scale-resonances in the
spectral domain $|m|,|n| \le 1000.$ In our computational domain we
have found 230464 such resonances, among them 213760 collinear (i.e.
all 4 wave vectors are collinear) and only 16704 (7.25\%)
non-collinear resonances. However, the nonlinear interaction
coefficient in collinear quartets is equal to zero and, therefore,
they have no dynamical significance \cite{zakh}. On the other hand,
for mathematical completeness we will consider these solutions too,
 because
 there may exist  4-wave  systems with the same
 dispersion law but with different forms of interaction coefficients
such that  are not necessarily zero on the collinear quartets.
\begin{figure}[h]
\begin{center}
\includegraphics[width=4cm]{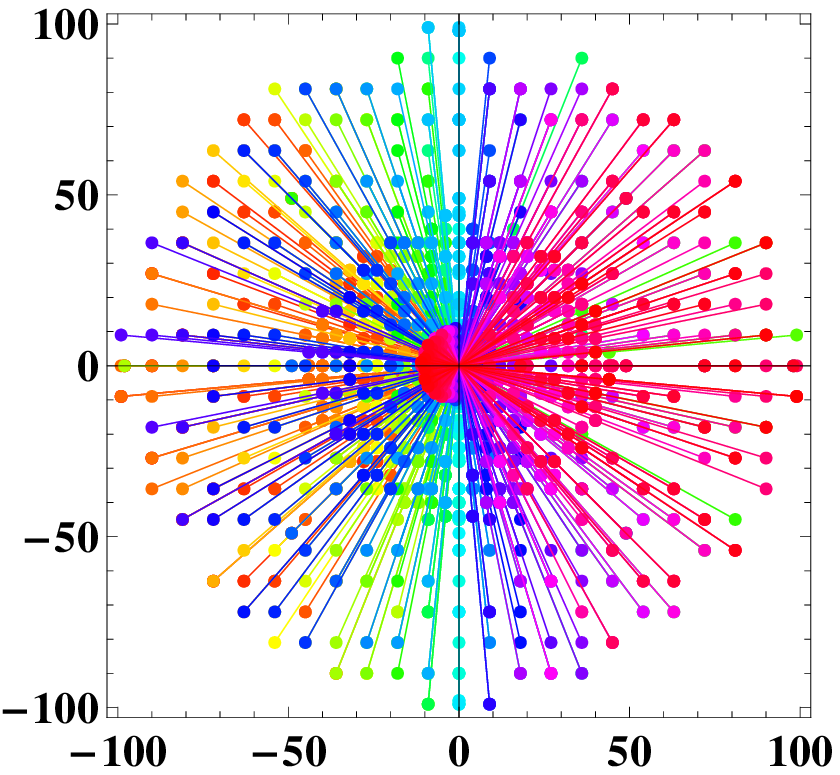}
\includegraphics[width=4cm]{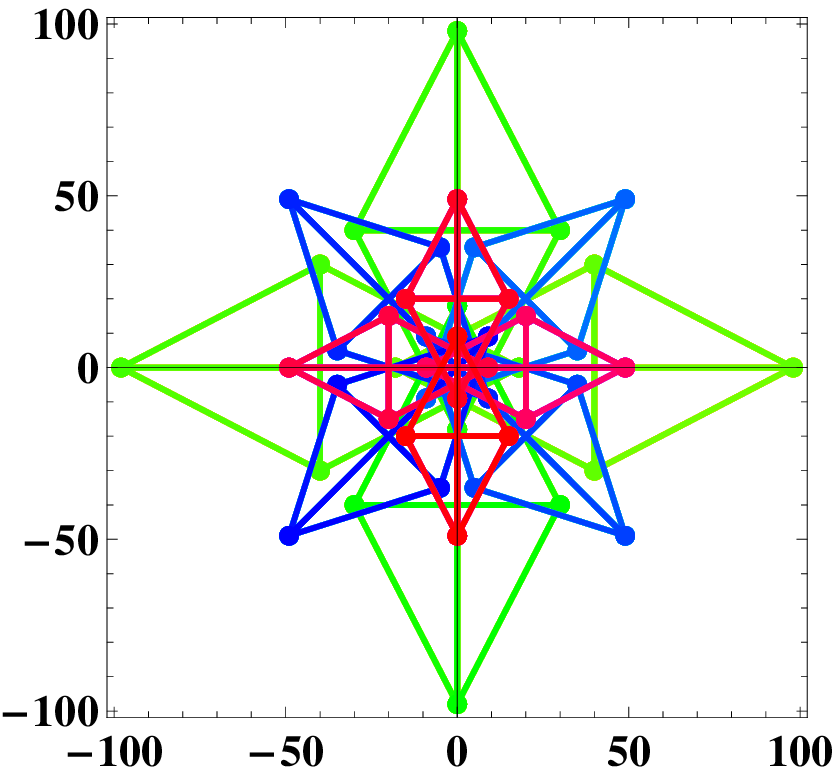}
\end{center}
\caption{Color online. Structure of collinear resonances in the
spectral domain $|{\bf k}|\le 100$ (left panel) and of non-collinear
resonances in the same domain (right panel) } %%
\label{f:ColNonCol}
\end{figure}

 In the Table I the structure of
all clusters is presented while in the Table II - the structure of
non-collinear is given; cluster length is the number of quartets
belonging to one cluster.

%%%\begin{table}[h]
%%%\begin{tabular}{|c|c|}
%%%  \hline
%%% % after \\: \hline or \cline{col1-col2} \cline{col3-col4} ...
%%%  Cluster length & Number of clusters \\ %%
%%%  \hline
%%% 4 & 43136 \\
%%% 8 & 1256 \\
%%% 48 & 452 \\
%%% 56 & 184 \\
%%% 64 & 20 \\
%%% 96 & 14 \\
%%% 112 & 10 \\
%%% 256 & 7 \\
%%% 304 & 2 \\
%%% 352 & 3 \\
%%% 368 & 6 \\
%%% 416 & 1 \\
%%% 512 & 1 \\
%%% 592 & 2 \\
%%% 800 & 1 \\
%%% 1024 & 1 \\
%%% 1152 & 1 \\
%%% 1376 & 1 \\
%%%  \hline
%%%\end{tabular}
%%%\caption{Clustering in the entire set of quartets in the domain
%%%1000x1000}
%%%\end{table}

%%
\begin{table}[h]
\begin{tabular}{|c|c||c|c|}
  \hline
  Cluster & Number of &   Cluster & Number of \\ %%
  length  & clusters  &   length  & clusters  \\
  \hline
 1 & 43136 & 56 & 3 \\
 2 & 1256 & 60 & 6 \\
 8 & 452 & 72 & 1 \\
 10 & 184 & 80 & 1 \\
 12 & 20 & 92 & 2 \\
 16 & 14 & 128 & 1 \\
 20 & 10 & 152 & 1 \\
 40 & 7 & 176 & 1 \\
 48 & 2 & 208 & 1 \\
  \hline
\end{tabular}
\caption{Clustering in the entire set of quartets in the domain
1000x1000 (symmetrical solutions are omitted)}
\end{table}

\begin{table}[h]
\begin{tabular}{|c|c||c|c|}
  \hline
  Cluster & Number of &   Cluster & Number of \\ %%
  length  & clusters  &   length  & clusters  \\
  \hline
 1 & 1312 &  10 & 18 \\
 2 & 48 &  12 & 6 \\
 3 & 8 &  16 & 8 \\
 6 & 14 &  34 & 2 \\
 8 & 4 &  46 & 2 \\
  \hline
\end{tabular}
\caption{Clustering in the \emph{non-collinear} subset of quartets
in the domain 1000x1000 (symmetrical solutions are omitted)}
\end{table}

%%%\begin{table}[h]
%%%\begin{tabular}{|c|c|}
%%%  \hline
%%%  % after \\: \hline or \cline{col1-col2} \cline{col3-col4} ...
%%% Cluster length & Number of clusters \\ %%
%%%  \hline
%%% 8 & 1312 \\
%%% 16 & 48 \\
%%% 24 & 8 \\
%%% 48 & 14 \\
%%% 64 & 4 \\
%%% 80 & 18 \\
%%% 96 & 6 \\
%%% 128 & 8 \\
%%% 272 & 2 \\
%%% 368 & 2 \\
%%%  \hline
%%%\end{tabular}
%%%\caption{Clustering in the \emph{non-collinear} subset of quartets
%%%in the domain 1000x1000.}
%%%\end{table}

 In Fig.~\ref{f:ColNonCol}
structure of collinear and non-collinear quartets is shown ia a
smaller domain $\sqrt{m_i^2 + n_i^2} \le 100.$

\subsection{Collinear quartets} First of all, let us make an important remark.
If a 4-tuple $ \{\{m_1,n_1\}, \{m_2,n_2\},\{m_3,n_3\},\{m_4,n_4\}\}$
consists of all collinear wave vectors, then the ratio $|m_i|/|n_i|$
is the same for all 4 wave vectors. Let us assume that $m_i, \ n_i
\ne 0$, then%%
\be \label{collinear} 0 \ne |n_i|/|m_i| = c,\   \ \ \forall
i=1,2,3,4\,, \ee %%
where $c$ is an arbitrary finite rational
 constant and%%
\be \label{transform}|{\bf k}_i|^{1/2}
=(m_i^2+n_i^2)^{1/4}=m_i^{1/2}(1+c^2)^{1/4}\,,\ee%%
 and Sys.(\ref{4grav}) takes the
form%%
 \bea\label{identity} \nonumber
\begin{cases}\nonumber
{m_1}^{1/2} + {m_2}^{1/2}={m_3}^{1/2}+{m_4}^{1/2}\,, \\
m_1+m_2=m_3+m_4\,,
\end{cases}
\RA \\
\begin{cases}\nonumber
(m_1m_2)^{1/2}=(m_3m_4)^{1/2}
\end{cases} \RA |m_1|=|m_3 m_4|/|m_2|.
 \eea%%
Now we can compute $m_1$ taking arbitrary integer $m_2, m_3, m_4$
provided that $|m_3 m_4|$ is divisible on $|m_2|$, and keeping in
mind that $n_i=c m_i$, we can find all collinear solutions with $c
\ne 0.$ Obviously, a rational number $c$ defines a line in the
spectral space and not all the lines are allowed. Case $c=0$
corresponds to the solutions lying on the axes X ($n_i=c m_i$) and Y
($m_i=c n_i$), i.e. with all $m_i=0$ or $n_i=0$ correspondingly,
e.g. 4-tuple $\{\{4,0\}, \{-49,0\}, \{-36,0\}, \{-9,0\}\}$.
Parametrization of the resonances in this case was first given in
\cite{zakh}. As we have already mentioned before, these quartets are
dynamically irrelevant because there is no nonlinear interaction
within these quartets \cite{zakh}.

\subsection{Non-collinear quartets}

\begin{figure}
  \begin{center}
    \includegraphics[width=4cm]{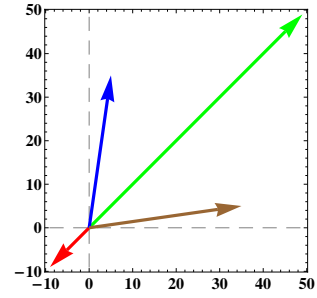}
  \end{center}
  \caption{Color online. First non-axial trident}
  \label{f:1stTrident}
\end{figure}
\subsubsection{Tridents}
Non-collinear scale-resonances have been studied first in
\cite{Naz-2006,LNP-2006} in the spectral domain $\sqrt{m_i^2+ n_i^2}
\le 100$ and a special type of quartets named {\it tridents} have
been
singled out. By definition, a wave quartet is called a trident if:\\
1) There exist two vectors among four in a quartet, say $\textbf{k}_1$ and $\textbf{k}_2$,
such that $\textbf{k}_1 \uparrow \! \downarrow \textbf{k}_2$, thus they satisfy\\ $(\textbf{k}_1\cdot \textbf{k}_2) = - k_1 k_2\,$;\\
2) Two other vectors in the quartet, $\textbf{k}_3$ and $\textbf{k}_4$, have the same length: $k_3 = k_4\,$;\\
3) $\textbf{k}_3$ and $\textbf{k}_4$ are equally inclined to
$\textbf{k}_1$, thus\\ $(\textbf{k}_1\cdot \textbf{k}_3) =
(\textbf{k}_1\cdot \textbf{k}_4)\,$.

%Namely, wave vectors ${\bf k}_1, \ {\bf k}_2$ are parallel and oppositely directed while $|{\bf k}_3|=|{\bf k}_4|$ and have the same angle with ${\bf k}_1$.

The following presentation for
a trident quartet has been suggested in \cite{Naz-2006,LNP-2006},
 \be \label{trident}
{\bf k}_1=(a,0),\  {\bf k}_2=(-b,0),\  {\bf k}_3=(c,d),\ {\bf
k}_4=(c,-d)\,, \ee%%

 and  two-parametric series of solutions has
been written out:%%
\be \label{ser} \begin{cases}a=(s^2+t^2+st)^2, \ b=(s^2+t^2-st)^2, \\
c=2st(s^2+t^2), \ d=s^4-t^4\,, \end{cases}\ee%%
with arbitrary integer $s,t$. It is easy to check that vectors $ \
{\bf k}_1, \ {\bf k}_2, \ {\bf k}_3, \ {\bf k}_4 \ $ belong to the
same class $Cl_1$, with weights
$$
\g_1=s^2+t^2+st,\ \  \g_2=s^2+t^2-st,\ \  \g_3=\g_4=s^2+t^2,
$$
and obviously $\g_1+\g_2=\g_3+\g_4, \ \forall s,t \in \Integer.$

Parametrization (\ref{trident}) corresponds to the tridents oriented
along the X-axis with its vectors ${\bf k}_1$ and  ${\bf k}_2$ and,
therefore, we will call them {\it axial} tridents. There exist also
{\it non-axial} tridents, for instance, the quartet
$\{\{49,49\},\{-9,-9\},\{5,35\},\{35,5\}\}$. As mentioned in
\cite{Naz-2006,LNP-2006}, all the { non-axial} tridents can be
obtained from the axial ones, (\ref{trident}), via a rotation by
angles with rational values of cosine combined with respective
re-scaling (to obtain an integer-valued solution out of
rational-valued ones).

\begin{table}[h]
  \begin{tabular}{|c|c|}
    \hline
    Cluster length & Number of clusters \\ %%
    \hline
      1 & 1320 \\
      2 & 48 \\
      6 & 40 \\
      12 & 10 \\
      18 & 2 \\
      22 & 2 \\
    \hline
  \end{tabular}
  \caption{Clustering in the subset of \emph{tridents} in the domain 1000x1000 (symmetrical solutions are omitted)}
\end{table}

In the computational domain $|m|,|n| \le 1000$, we have found 13888
non-axial tridents, the first non-axial trident is shown on the
Fig.~\ref{f:1stTrident}. Among 13888, 13504 tridents have no vectors
on any axis and 384 tridents have just one vector on an axis (for
instance, $\{\{180,135\},\{0,64\},\{120,119\},\{60,80\}\}$). The
total amount of all possible tridents is 14848, thus only 960
are axial ($6.5\%$ of the total number).
 The data on tridents' clustering are given
in the Table III.

\subsubsection{Non-tridents}

 Our study of the resonance solution set in the spectral domain
$|m_i|,|n_i|\le1000$ shows that not all non-collinear cascading
quartets are \emph{tridents}. They are called further
\emph{non-tridents}, e.g. a quartet $\{\{990,180\}, \{128, 256\},
\{718, 236\}, \{400, 200\}\}$ is a non-trident quartet.

\begin{figure}[h]
\begin{center}
\includegraphics[width=4cm]{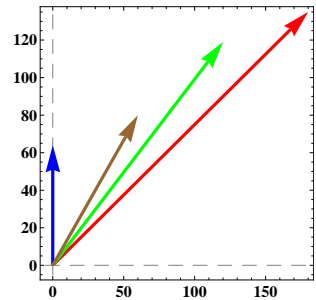}
\end{center}
\caption{Color online. First non-trident} \label{f:skew-quartets}
\end{figure}

The overall number of tridents is 14848 while the number of
non-tridents is 1856. Notice that the first non-trident quartet $
\{\{180,135\}, \{0,64\}, \{120,119\}, \{60,80\}\} $ lies in the
spectral domain $|{\bf k}|\le 225$. This means, that if we are
interested only in the large-scale quartets, say, quartets with
$|{\bf k}|\le 100,$ the complete set of scale-resonances consists of
1728 quartets, among them - 1632 collinear quartets and 96 tridents,
but no non-tridents yet.

\subsection{Clusters}

\begin{figure}[h]
  \begin{center}
    \includegraphics[width=4cm]{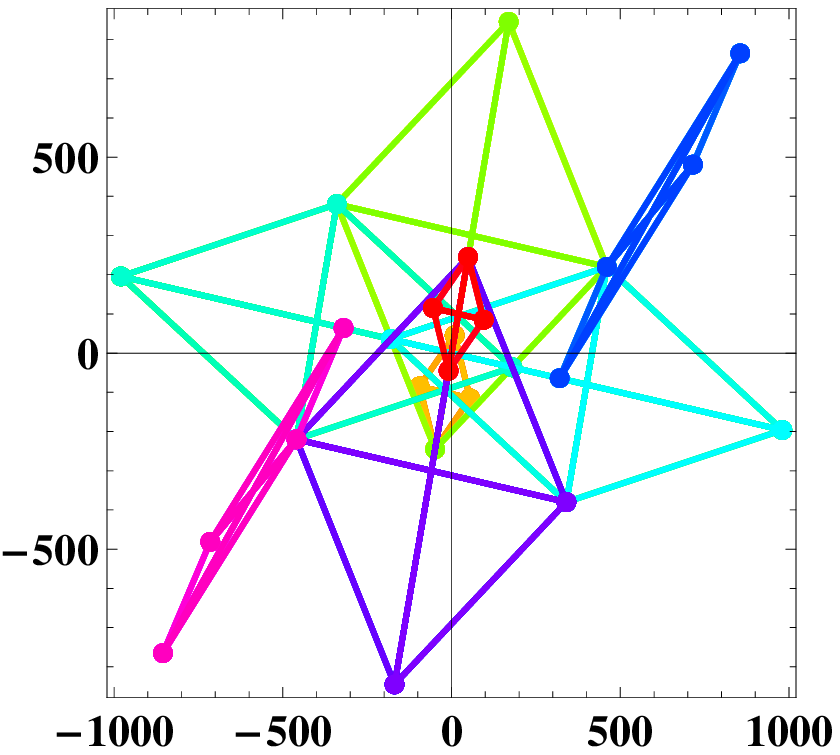}
    \includegraphics[width=4cm]{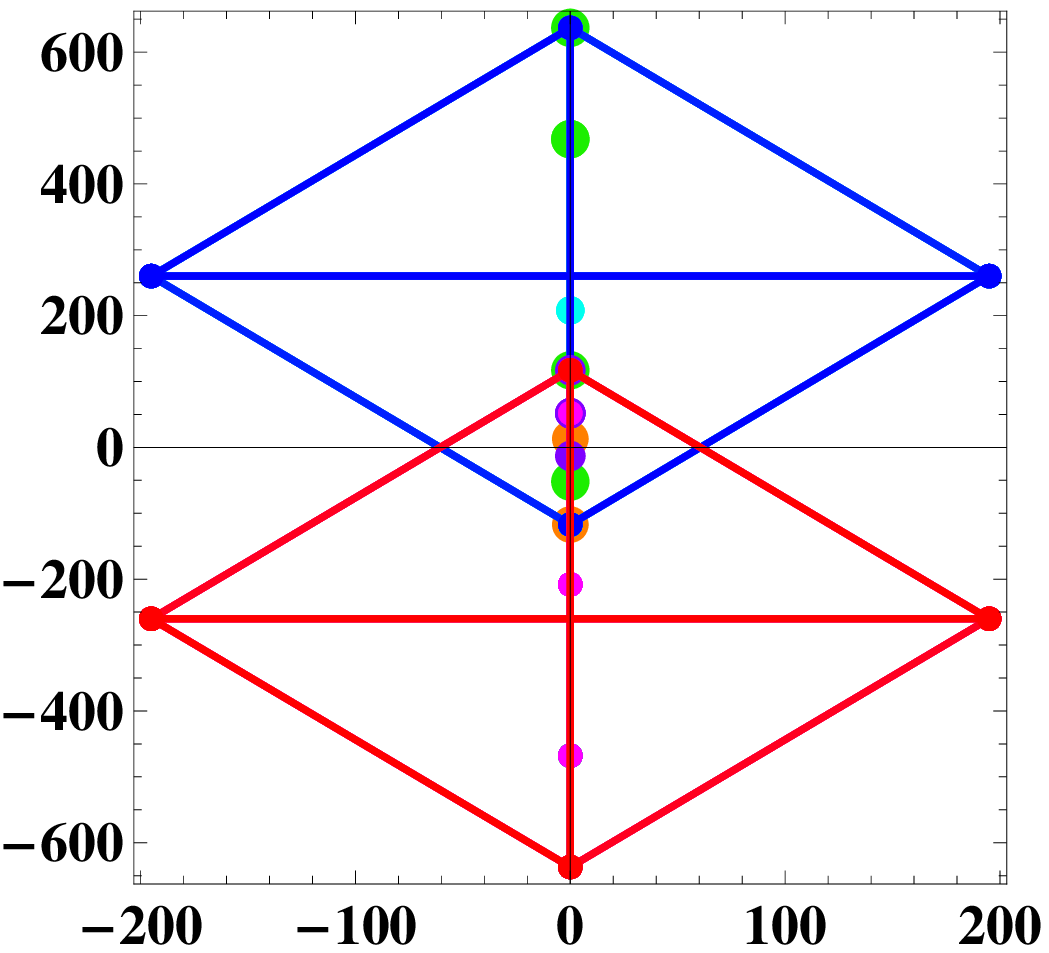}
  \end{center}
\caption{\label{f:Ru01} Color on line. {\bf Left panel}: A shortest
cluster formed by both tridents and non-tridents; cluster length is
8 (64 with 8 symmetries), among them 6 (48 with 8 symmetries)
tridents and 2 (16 with symmetries) non-tridents. {\bf Right panel}:
A shortest cluster formed by both collinear and non-collinear
quartets; cluster length is 8 (48 with 8 and 4 symmetries), among
them 6 (32 with 8 and 4 symmetries ) collinear and 2 (16 with 8
symmetries) non-collinear quartets. No multi-edges are shown.}
\end{figure}

In the previous Section we have shown that there are three different
types of scale-resonances: collinear quartets, tridents and
non-tridents. There are also clusters formed by different types of
scale-resonances, for instance, clusters containing tridents and
non-tridents (Fig.~\ref{f:Ru01}, left panel) or collinear and
non-collinear quartets  (Fig.~\ref{f:Ru01}, right panel). Notice
that,  since collinear quartets of gravity water waves have zero
interaction coefficients~\cite{zakh},  the cluster  shown on the
right panel can dynamically be regarded as two independent quartets.

%%

One of the most important characteristics of the resonance structure
is the {\it wavevector multiplicity} (introduced in \cite{KK06-2}),
which describes how many times a given wavevector is a part of some
solution. In the Table IV the wavevectors multiplicities are given
for non-collinear quartets. It turned out that $91\%$ of all these
wavevectors (6720 from overall amount 7384) have multiplicity {1
(counting the 8
 symmetry generated solutions as the same quartet.)

\begin{table}[h]
  \begin{tabular}{|c|c|}
    \hline
     \text{Multiplicity} & \text{Amount of vectors} \\ %%
    \hline
 1 & 6720 \\
 2 & 424 \\
 3 & 192 \\
 4 & 36 \\
 5 & 8 \\
 6 & 4 \\
    \hline
  \end{tabular}
  \caption{Wavevectors multiplicity computed for the non-collinear quartets in the spectral
  domain $|m_i|,|n_i|\le 1000$ (symmetrical solutions are omitted)}
\end{table}

 In Fig.~\ref{f:smallest}, we see
an example of such a simple cluster consisting from just one
physical quartet. In the Fig.~\ref{f:skew-quartets} the cluster of 4
connected quartets is shown (the symmetry generated solutions are
omitted), with all together only 7 different wave-frequencies.

\begin{figure}[h]
  \begin{center}
    \includegraphics[width=7cm]{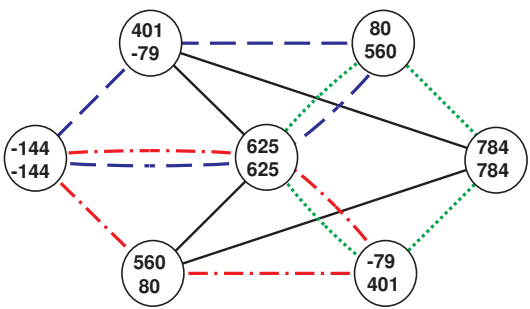}
  \end{center}
  \caption{Color on line. Example of a non-trident cluster of length 4.}
  \label{f:skew-quartets}
\end{figure}

Obviously, the cluster structure defines the form of dynamical
system corresponding to the cluster. The main motivation of our
detailed study of clusters  is, of course,  in constructing an
isomorphism (i.e. one-to-one correspondence) between
 a cluster and a dynamical system. In \cite{KM07} this construction
 has been presented for an arbitrary 3-wave resonance system, with
 triads as primary elements of the planar graph.
 Its implementation in Mathematica was given in \cite{all08}, where
  interaction coefficients similar to
$Z$ in (\ref{4grav-dynamics}) were also computed. To construct this
isomorphism two following facts were used: 1) in a 3-wave
resonance system only {\it scale-resonances}  exist, and 2) if we
add an arbitrary triad to a cluster, in general we always add some
new wave-frequencies as well (the only exception is identified in
\cite{KM07}). A 4-wave resonance system does not possess these nice
properties; on the contrary - there exist scale- and
angle-resonances, and most of the quartets are parts of  symmetry
generated solution sets. It would be a challenge to develop a
general approach to construct dynamical systems for resonance
clusters in an arbitrary 4-wave system.

\section{Angle-resonances}

Until now we have studied the structure of the scale-resonances
which are described by (\ref{1class}) and
 which generate new scales (i.e. new
values of $|k|$, thereby providing the energy cascade mechanism. The
angle-resonances represented by (\ref{2class}) do not generate new
scales but  they can redistribute energy among the modes which were
excited initially (or which are forced externally).
 Taken on their own,
these resonances could lead to the thermal equilibrium distribution
of the energy and the waveaction among the (initially excited)
resonant waves, if the number of such waves is large, or they could
lead to a periodical behavior or a strange attractor, if the number
of the initially excited modes is small.
 The
important fact, however, is that the scale-  and the
angle-resonances are not independent and can form a {\it mixed
cluster} containing both types of this resonances \cite{K07}.
 The energy cascade mechanism in such a mixed cluster is presented
schematically in Fig.\ref{f:MixedCascade} where the quadrangles $\ S_1 \
$ and $\ S_2 \ $ denote scale-resonances so that 4-tuples $\
(V_{1,1},...,V_{1,4} )$ and $\ (V_{2,1},...,V_{2,4} )$  represent
scale-resonances and squares $\ A_1,... A_i,.., A_n \ $ represent
angle-resonances.%%
\begin{figure}[h]
\begin{center}
\includegraphics[width=8cm,height=4cm]{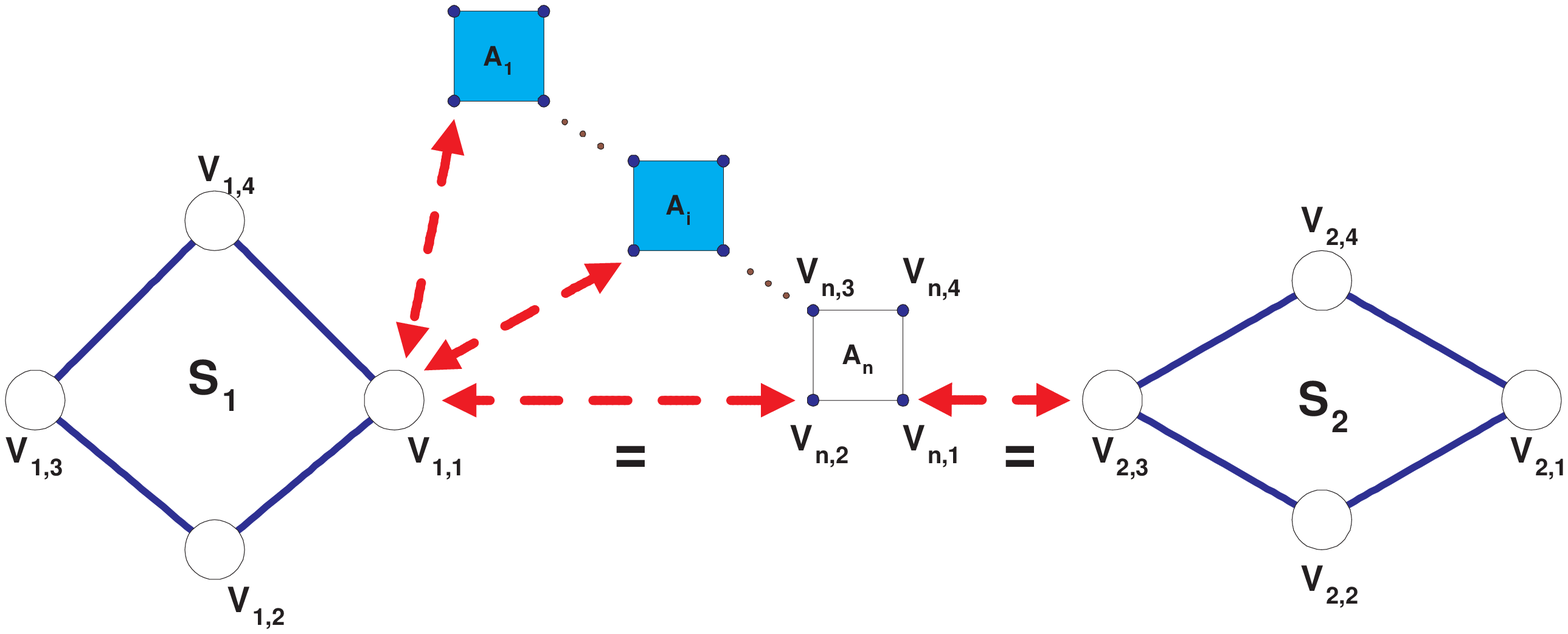}
\end{center}
\caption{\label{f:MixedCascade} Color on line. Schematic
presentation of mixed resonance cascade}
\end{figure}%%

If a wave takes part simultaneously in angle- and scale-resonances,
corresponding nodes of $\ S_1, S_2 \ $ and $\ A_j \ $ are connected
by (red) dashed arrows so that $\ V_{1,1}=V_{n,2}\ $ and $\
V_{n,1}=V_{2,3}. \ $ We have found many examples of such mixed
clusters
%this type of resonances - called further {\it mixed} resonances -
in our solution
set, for instance:%
 $$
 S_1=\{\{-64,-16\},\{784,196\},\{144, 36\},\{576, 144\}\},$$%
$$ A_n= \{\{ -64, -16\},\{4,16\},\{-64,16\},\{4 ,-16)\}\},$$%
 with $ V_{1,1}=V_{n,2}=( -64, -16),$
 $$S_2=
 \{\{-49,-196\},\{4,16\},\{-36,-144\},\{-9, -36\}\},$$%
which is further on connected with an angle-resonance
$$ A_{\tilde{n}}=
\{\{-49,-196\},\{784,196\},\{-49,196\},\{784,-196\}\}, \ $$%
(not shown in Fig.\ref{f:MixedCascade}) {\it via} a node-vector
$(-49,-196)$.

The energy flux over scales due to the mixed cluster is weak: one wave
can participate in a few dozen of angle-resonances (see Fig.
\ref{f:weak}), which means that only a small part of its energy will go
to a scale-resonance. Moreover, even this weak energy cascade may
terminate at finite wavenumber if it happens to move along a {\em
finite} cluster which terminates before reaching the dissipation
range (or before reaching the range of high frequencies where
the nonlinearity gets large enough for the quasi-resonances to take over
the energy flux from the exact resonances).

\begin{figure}[h]
\begin{center}
\includegraphics[width=8cm,height=4cm]{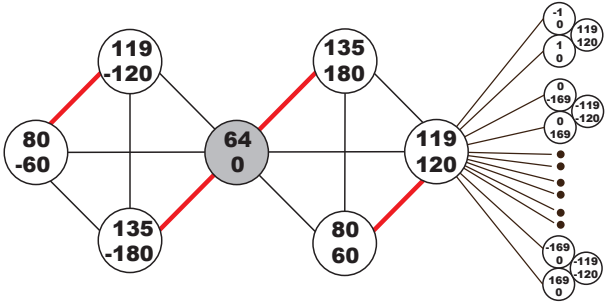}
\end{center}
\caption{\label{f:weak} Color on line. Wave (64,0) takes part in 2
scale-resonances, both {\it non-tridents}. The upper number in the
circle is $m$, the lower is $n$, and (red) thick lines drawn between
vectors on the same side of the Eqs.(\ref{4grav}). Wave (119,120)
takes part in 1 scale-resonance and in 12 angle-resonances.}
\end{figure}%%
The number of angle-resonances in the spectral domain $|m_i|, |n_i|
\le 1000$ is of order of $6 \cdot10^8$  while the number of
scale-resonances in the same domain is of order of $8\cdot10^5$,
among them less than $2\cdot10^4$ are non-collinear and do play a
role in the energy exchange among the modes within the quartets (see
next the Section).  It is easy to see that an arbitrary wavevector
$(m,n)$ takes part in infinite number of resonances if spectral
domain is unbounded. Indeed, let us fix $m$ and $n,$ then a
quartet%
 \be \label{Angle1}
 (m,n)(t,-n)\rightarrow (m,-n)(t,n)
 \ee%%
a scale-resonance with arbitrary $t=0,\pm 1, \pm2, \ldots$. More
involved 5-parametric series of angle-resonances found from the
following considerations. For angle-resonances of four wavevectors
$(a,b)(c,d)\rightarrow(p,q)(l,m)$, Eqs.(\ref{4grav}) can be
rewritten as%%
 \bea \label{Angle2}
 \begin{cases}
 a^2+b^2=p^2+q^2, \quad c^2+d^2=l^2+m^2\\
 a+c=p+l,\quad b+d= q+m
 \end{cases}
\eea%
Simple algebraic transformations and known parameterizations of the
sum of two integer squares (e.g. for a circle or for the Pythagorean
triples) yield
 \bea\label{Angle3}
 \begin{cases}
a=(s^2-t^2)/(s^2+t^2), \quad b=2st/(s^2+t^2),\\

p=(f^2-g^2)/(f^2+g^2),\quad q=2fg/(f^2+g^2), \\
d =(a^2+b^2+ac-ap-cp-bq)/(q-b)
 \end{cases}
\eea%
This is an easy task to check then that the solutions of
(\ref{Angle2}) can be written out (perhaps with  repetitions)
\emph{via} five integer parameters $s, \ t, \ f, \ g, \ c$ (rational
solutions should be renormalized to integer).
Notice that (\ref{Angle1}) degenerates to  trivial resonances%
 \be \label{Angle4}
 (m,0)(t,0)\rightarrow(m,0)(t,0),
 \ee%%
  if $n=0$, i.e. it does not include
any resonances of wavevectors of the form $(m,0)$, for instance
$(1,0)$. In this case the choice $f=1, \ g=1$ in (\ref{Angle3})
gives%
 \be \label{Angle5}
 (1,0)(c,1+c)\rightarrow (0,1)(1+c,c).
 \ee%%
 Analytical series are very
helpful not only for computing resonance quartets and clusters
structure but also while investigating the asymptotic behavior of
interaction coefficients.

\begin{figure}[h] \centerline{\psfig{file=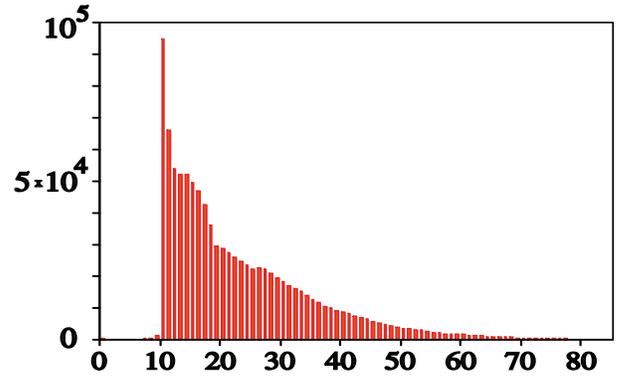,
width=8cm,height=5cm}} \vspace*{8pt} \caption{The multiplicities
histogram for angle-resonances. \label{f:VecMul222}}
\end{figure}%%
The multiplicity histogram for the angle-resonances is shown in
Fig.\ref{f:VecMul222}. On the axis $X$ the multiplicity of a vector
is shown and on the axis $Y$ the number of vectors with a given
multiplicity. This  graph has been cut off - multiplicities go very
high, indeed the vector (1000,1000) takes part in 11075 solutions.
All this indicates that the angle-resonances play an important role
in the overall dynamics of the wave field.

\section{Dynamics}\label{s:dyn}

\subsection{The wave field}
Once the clusters are found, one can consider an evolution of
amplitudes of waves that belong to each individual cluster by
considering a respective reduction of the dynamical equation. For the
gravity wave case, the appropriate dynamical equation is Zakharov
equation for the complex amplitude $a_{\bf k}(t)$ corresponding to
the ${\bf k}$-mode, \be i {da_{\bf k} \over dt} = \sum_{{\bf k_1},
{\bf k_2},{\bf k_3}} ^{{\bf k}+{\bf k_1} = {\bf k_2}+{\bf k_3}}
 T^{{\bf k},{\bf k_1}}_{{\bf k_2},{\bf k_3}} a^*_{\bf k_1} a_{\bf k_2}
a_{\bf k_3} e^{i (\omega_{\bf k}+\omega_{\bf k_1} - \omega_{\bf
k_2}- \omega_{\bf k_3}) t}, \ee where $T^{{\bf k},{\bf k_1}}_{{\bf
k_2},{\bf k_3}} \equiv T({\bf k},{\bf k_1}, {\bf k_2},{\bf k_3})$ is
an interaction coefficient for gravity water waves which can be
found in \cite{dynamics2}.

For very weak waves, amplitudes $a_{\bf k}(t)$ vary in
time much slower than the linear oscillations.
The factor $e^{i (\omega_{\bf k}+\omega_{\bf k_1} -
\omega_{\bf k_2}- \omega_{\bf k_3}) t}$ on RHS
will rapidly oscillate for most waves except for
those in an exact resonance for which this factor is 1.
Thus, only the resonant modes will give a contribution
to the dynamics in this case and the oscillating contributions
of the non-resonant terms will average out in time to zero and
will not give any contribution to the cumulative change
of $a_{\bf k}(t)$.
Leaving only the resonant terms, we have
\be
i {da_{\bf k} \over dt} = \sum^R_{{\bf k_1}, {\bf k_2},{\bf k_3}}
 T^{{\bf k},{\bf k_1}}_{{\bf k_2},{\bf k_3}} a^*_{\bf k_1} a_{\bf k_2}
a_{\bf k_3} , \label{cluster-dyn} \ee where $\sum^R $ means
a summation only over ${\bf k_1}, {\bf k_2}$ and ${\bf k_3}$ which are
in resonance with ${\bf k}$. Obviously, we should consider ${\bf
k}$'s from the same cluster only (i.e. solve the problem for one
cluster at a time). Note that the fast timescale of the linear
dynamics completely disappeared from this equation. Thus,
paradoxically, the dynamics of very weak waves in finite boxes is
strongly nonlinear: it is more nonlinear than in WTT which works for
larger amplitudes and where quasi-resonances ensure phase
randomness. This explains the fact found in the three-wave example
that  even relatively large clusters often exhibit a periodic or
quasi-periodic behavior \cite{KL-2008}.

A study of the system (\ref{cluster-dyn}) is  possible analytically
for small clusters (perhaps even integrating the system in some lucky cases)
and numerically for large clusters, which is an interesting subject for future
research. Some properties, however, can already be seen in the example of a single
quartet which has been studied before \cite{dynamics1}.
Let us briefly discuss these properties.

\subsection{A quartet}

The dynamical system describing slowly changing amplitudes of a quartet
has the form \cite{dynamics1}:%%
\bea \label{4grav-dynamics}
\begin{cases}
i \dot{a}_1= 2Z a_2^*a_3a_4\\
i \dot{a}_2= 2Z a_1^*a_3a_4\\
i \dot{a}_3= 2 Z^* a_4^*a_1a_2\\
i \dot{a}_4= 2 Z^* a_3^*a_1a_2
\end{cases}
\eea %%
and  $Z =  T^{{\bf k_1},{\bf k_2}}_{{\bf k_3},{\bf k_4}}$. The
mathematical analysis of the system (\ref{4grav-dynamics}) can be
performed similar to what has been done in \cite{KL-06} for an
integrable 3-wave system of resonantly interacting planetary waves,
though computations of the modulus of elliptic integral are more
involved and variety of different dynamical scenarios is
substantially reacher (see \cite{dynamics1} for details). The
general answer can be given in terms of Jacobean elliptic functions.

Obviously, the quartets with interaction coefficient $Z=0$ do not
influence the general dynamics of the wave field at the corresponding
time scale. As it was shown in \cite{zakh},
 for all {\it collinear} quartets $Z \equiv 0$,
and therefore they can be excluded form consideration.

On the other hand,  in \cite{dynamics3} so-called "degenerate
quartets" ({\it tridents} in our terminology) have been studied and
it was established numerically that they have strictly periodic
behavior: after appropriate translation in the horizontal plane, two
snapshots of the free-surface taken at $t=0$ and $t=T$ are
identical.

For a general quartet and for arbitrary and arbitrary initial
conditions, the system (\ref{4grav-dynamics}) "does not exhibit
strict periodicity" in numerical simulations \cite{dynamics1}. Thus,
the general formulae for the solutions of (\ref{4grav-dynamics})
have to be studied in more details in order to distinguish periodic
and non-periodic dynamics of an arbitrary quartet. On the other
hand, since our main interest in this paper is the large-scale
dynamics, all scale-resonances with a non-zero interaction
coefficient are tridents and therefore demonstrate a periodic time
behavior. Of course, if a trident  is involved into a cluster with
some other quartets then one should be cautious about the
predictions obtained for an isolated quartet.

\section{Summary and Discussion of Results}

\textbullet~In this paper, we studied properties of deep water
gravity waves bounded by a square periodic box. At very small wave
amplitudes, when the nonlinear resonance broadening is less than the
$k$-space spacing, WTT fails and only the waves which are in {\em
exact} four-wave resonance can interact. This situations appears to
be typical for all existing numerical simulations \cite{LNP-2006}
and laboratory experiments \cite{denis}. Thus, to understand the
wave behavior in laboratory experiments and in numerical simulations
it is crucial to study exact resonances among discrete wave modes,
which was the focus of the present paper. Of course, the notation of
"very small amplitudes"  has to be worked out explicitly for
interpreting the results of laboratory experiments. The smallness of
amplitudes is defined by the choice of a small parameter $0< \e \ll
1$ and depends of the intrinsic characteristics of the wave system,
for instance, for atmospheric planetary waves it is usually taken as
the ratio of the particle velocity to the phase velocity which
allows to obtain explicit estimation \cite{kar2} for a wave
amplitude $a(m,n)$ (corresponding to the weakly nonlinear regime) as
a function of $m$ and $n$. For the water surface waves, the wave
steepness, $|a(m,n)|/(m^2+n^2)^{1/2} L $, is usually taken as a
small parameter $\e$ and  $\e \sim 0.1$ corresponds then to the
weakly nonlinear regime. { For such weakly nonlinear waves to feel
discreteness of the $k$-space, their nonlinear frequency broadening
has to be less than the distance between adjacent $k$-modes. In
terms of the wave steepness this condition reads $\e <
(m^2+n^2)^{-1/8}$, see \cite{Naz-2006,LNP-2006}.

\textbullet~We found numerically all resonant quartets on the set of
1000 x 1000 modes making use of the $q$-class method originally
developed in \cite{K06-3}. We found that all resonant quartets
separate into noninteracting with each other clusters.  Each cluster
may consist of two types of quartets: scale- and angle-resonances.
The angle-resonances cannot transfer energy to any $k$-modes which are
not already present in the system. They cannot carry an energy flux
through scales and their main role is to thermalize the initially
excited modes. The scale-resonances are much more rare than the angle
ones and yet their role is important because they are the only
resonances that can transfer energy between different scales. Most
of the scale-resonances, but not all, are of the {\it trident} type,
for which a partial parametrization can be written out explicitly.
If one is interested in large-scale modes only, say 100 x 100
domain, then the tridents are the only scale-resonances, which is
very fortunate due to the available parametrization.

\textbullet~Even though the angle-resonances cannot cascade energy,
they are important for the overall cascade process because they are
involved in the same wave clusters with the scale-resonances. One
wave mode may typically participate in many angle resonances and
only one scale resonance. Thus, one can split large clusters into
"reservoirs", each formed by a large number of angle quartets in
quasi-thermal equilibrium, and which are connected with each other
by sparse links formed by scale quartets. This structure suggests a
significant energy cascade slowdown and anisotropy with respect to
the infinite-box limit. Further study is needed to examine the
structure of such large clusters and possible energy cascade routes
from the region of excitation at low wavenumbers to the dissipative
large $k$ range. In the essentially finite domain the situation is
opposite.  Quite recently results of the laboratory experiments with
surface waves on deep water were reported \cite{SH07,CHOS06,HPS06}
in which regular, nearly permanent patterns of the water surface
have been observed. A feasible way to interpret these results would
be 1) to establish that the conditions of the experiments correspond
to the weakly nonlinear regime; if yes - to proceed as follows: 2)
to compute all exact resonances in the wave-lengths range
corresponding to those in the experiments; 3) to demonstrate that
for chosen wave-lengths and the size of laboratory tank no
scale-resonances appear, 4) to attribute the regular patterns to the
corresponding angle-resonances. Obviously, the scale-resonances
would produce the spectrum anisotropy and disturb the regular
patterns.

\textbullet~We also discussed consequences of the cluster structures
for the dynamics, and argued that one should expect a less random and
more regular behavior in the case of very low amplitude waves with
respect to larger (but still weak) waves described by WTT. More
study is needed in future both analytically, for small clusters, and
numerically, for large clusters. A particularly interesting question
to answer in this case is about any possible universal mechanisms of
transition between the regular dynamics to chaos and possible
coexistence of the regular and chaotic motions.

 \textbullet~ Last not least. The knowledge of the
2\textbf{D}-resonance structure might yield new insights into the
origin of some well-known physical phenomena, for instance,
Benjamin-Feir (B-F)instability~\cite{BF67} or McLean instability.
This is "a modulational instability in which a uniform train of
oscillatory waves of finite amplitude losses energy to a small
perturbation of waves with nearly the same frequency and
direction"~\cite{StabilBF}. As it was shown recently
in~\cite{HH03,HHS03,StabilBF}, the modulational instability, though
well established not only water waves theory but also in plasmas and
optics, has to be seriously reconsidered. It turned out that 1) it
can be shown analytically that arbitrary small dissipation
stabilizes the B-F instability, and 2) results of laboratory
experiments show that B-F theory generally over-predicts the growth
rate. Moreover, the growth rate changes with the
time~\cite{StabilBF}. Some researchers state even that "... this
effect is far less significant than was believed and should be
disregarded"~\cite{LY77}.  The other way to treat the problem would
be to try and and explain the modulational instability through
non-collinear (that is, essentially two-dimensional) exact
resonances~\cite{H95}. Similar questions arise in  the study of
McLean instabilities defined by  the magnitudes of the water depth
on which surface waves are studied. For instance, as it was
demonstrated in~\cite{FK05}, in some regimes of shallow water the
instabilities are due to higher order resonances among 5 to 8 waves.
It would, therefore, be interesting to see how the B-F and McLean
instabilities are modified by the finite flume effects and the
corresponding discreteness of the wave resonances, and to see what
role this could have played in the past laboratory and numerical
experiments.  Results presented in our paper can be regarded as a
necessary first step for such an investigation,  and the future work
would involve application of the
% higher order resonances and
$q$-class method to computing the higher order resonances which may
be involved in modulational instabilities.

\section*{Acknowledgments}
{\bf Acknowledgements}.  E.K. acknowledges the support of the
Austrian Science Foundation (FWF) under project P20164-N18 "Discrete
resonances in nonlinear wave systems". O.R. and S.N. acknowledge the
support of the Transnational Access Programme at RISC-Linz, funded
by European Commission Framework 6 Programme for Integrated
Infrastructures Initiatives under the project SCIEnce (Contract No.
026133).  Authors are genuinely grateful to both anonymous Referees
whose suggestions made the form our paper more clear and led, in
particularly, to including the very important paragraph about
Benjamin-Fair and McLean instabilities. Authors express a special
gratitude to Victor L'vov for fruitful and stimulating discussions.

\end{document}